\documentclass[aps,amsmath,amssymb,showpacs,floatfix,superscriptaddress,twocolumn]{revtex4}
\usepackage{graphicx}
\usepackage{dcolumn}
\usepackage{bm}
\usepackage{hyperref} 
\usepackage{latexsym}
\usepackage{float}

\def\dw{d_{\text{w}}}
\def\tdw{\tilde d_{\text{w}}}
\def\df{d_{\text{f}}}
\def\tdf{\tilde d_{\text{f}}}

\def\dk{d_{\text{k}}}
\def\tdk{\tilde d_{\text{k}}}

\def\tzeta{\tilde\zeta}

\begin{document}
\title{Fractal and Transfractal Recursive Scale-Free Nets}   

\author{Hern\' an D. Rozenfeld}
\email{rozenfhd@clarkson.edu}
\affiliation{Department of Physics, Clarkson University, Potsdam NY 13699-5820, USA}
\author{Shlomo Havlin}
\email{havlin@ophir.ph.biu.ac.il}
\affiliation{Minerva Center and Department of Physics, Bar-Ilan University, Ramat-Gan 52900, Israel}
\author{Daniel ben-Avraham}
\email{benavraham@clarkson.edu}
\affiliation{Department of Physics, Clarkson University, Potsdam NY 13699-5820, USA}

\begin{abstract}
We explore the concepts of self-similarity, dimensionality, and (multi)scaling in a new family of recursive
scale-free nets that yield themselves to exact analysis through renormalization techniques.  All nets in
this family are self-similar and some are fractals --- possessing a finite fractal dimension --- while others
are {\it small world\/} (their diameter grows logarithmically with their size) and are infinite-dimensional.  We show how a useful measure of {\it transfinite\/} dimension may be defined and applied to the small world nets.
Concerning multiscaling, we show how first-passage time for diffusion and resistance between {\it hubs\/} (the most connected nodes) scale differently than for other nodes.  Despite the different scalings, the Einstein relation between diffusion and conductivity holds separately for hubs and nodes.  The transfinite exponents of small world nets 
obey Einstein relations analogous to those in fractal nets. 
\end{abstract}

\maketitle

\maketitle

\section{INTRODUCTION}

\begin{figure}[ht]
  \vspace*{0.cm}\includegraphics*[width=0.40\textwidth]{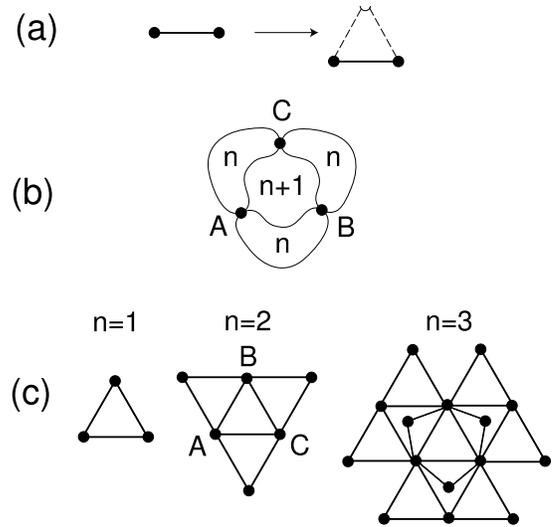}
\caption{The Dorogotsev-Goltsev-Mendes graph, or the $(1,2)$-flower.
(a)~First construction method: Each link in generation $n$ (solid line) is replaced by two parallel paths of $u=1$ (dotted line) and $v=2$ (broken lines) links long.  
(b)~A second method of construction that highlights self-similarity: Generation $n+1$ is obtained by adjoining three copies of generation $n$ at the hubs, denoted by A, B, and C.
(c)~Generations $n=1,2,3.$}
\label{DGM}
\end{figure}

Scale-free networks are ubiquitous in science and in everyday life, and have been the focus of intense interest~\cite{reviews}.  Recently, Song {\it et al\/}., have demonstrated that several naturally occurring scale-free networks exhibit {\it fractal\/} scaling~\cite{Song,Song2}.  
Previously, 
Dorogovtsev, Goltsev, and Mendes (DGM)~\cite{dorogovtsev} studied a {\it hierarchical\/}~\cite{hierarchical} scale-free net that is constructed
recursively  (Fig.~\ref{DGM}), in a manner reminiscent of exact fractal lattices such as the Sierpinski 
gasket~\cite{feder,bahbook}.  The DGM net is self-similar in a weak sense: it contains subgraphs that resemble the whole, but lacks the affine transformation of scale associated with self-similarity in fractals.  As a result, though resembling a fractal, the DGM net has infinite dimensionality  --- a fact that led Dorogovtsev {\it et al.},~\cite{dorogovtsev} to call it a {\it pseudofractal\/}.

In this paper we introduce study deterministic networks ---  $(u,v)$-{\it flowers\/} and 
$(u,v)$-{\it trees\/} --- that generalize the DGM net to a whole  family of scale-free nets,  of degree exponent $\gamma=1+\ln(u+v)/\ln2$. For $v\geq u>1$, networks in this family are self-similar, including an affine transformation of scale, and they posses well defined fractal dimensions.  For $u=1$
(the case including the DGM net), the networks are self-similar only in the weak sense, without the affine transformation, and they are infinite-dimensional.  We exploit their self-similarity to define {\it transfinite\/} dimensions: dimensionalities of ``higher cardinality" that usefully characterize the $(1,v)$-nets.  Accordingly,
we refer to the DGM and similar nets as {\it transfractals\/}.

Having fractal nets (for $u>1$), it is natural to wonder whether the useful lessons learned from regular fractals,
such as scaling relations between exponents, 
apply to them as well.  One reason to doubt that this might be the case is the broad range of node degrees in scale-free nets: this calls into question any result where self-averaging is invoked, that is, can nodes
of small and large degree be treated on an equal footing?  We explore these issues by focusing on the
Einstein relation between exponents for resistance, diffusion, and the fractal dimensions of our models.  We find that for fractal
nets the most connected nodes (hubs) scale in the same way as nodes of small degree, and that the Einstein relation is satisfied.  For transfractal nets, however, there are different scaling laws of resistance and diffusion for the hubs and for nodes of smaller degree.  We show that, nevertheless, the Einstein relation is satisfied, separately,
by each of the transfinite set of exponents characterizing  the two subsets of nodes.

\section{RECURSIVE SF FLOWERS AND TREES}

\begin{figure}[ht]
  \vspace*{0.3cm}\includegraphics*[width=0.40\textwidth]{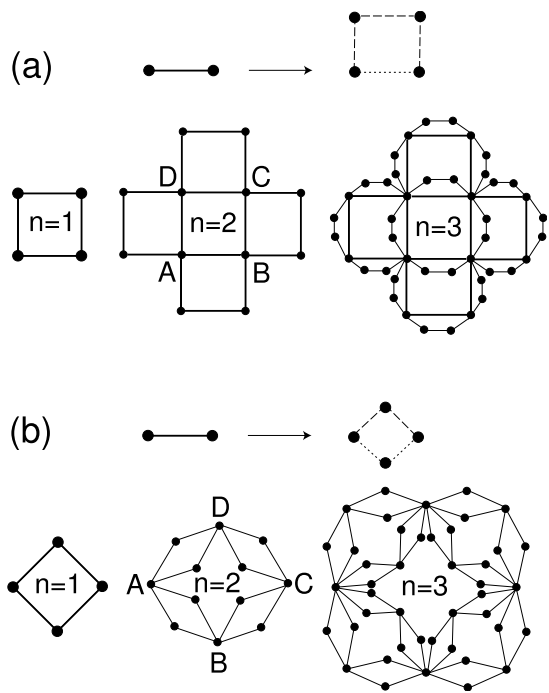}
\caption{$(u,v)$-flowers with $u+v=4$ ($\gamma=3$).
(a)~$u=1$ (dotted line) and $v=3$ (broken line). (b)~$u=2$ and $v=2$.  The graphs may also
be iterated by joining four replicas of generation $n$ at the hubs A and B, for (a), or A and C, for~(b).
}
\label{g3}
\end{figure}

We focus on a certain class of {\it hierarchical\/} nets~\cite{hierarchical}, that generalize the DGM net in the following way~\cite{other}.  Given a net of generation $n$, generation $n+1$ is obtained by replacing each link by two parallel paths of $u$ and $v$ links long.  A natural
choice for the genus at generation $n=1$ is a cycle graph (a ring) consisting of $u+v\equiv w$ links and nodes (other choices are possible).  In the following we assume that $u\leq v$, without loss of generality,
and we term a net obtained by this construction method a $(u,v)$-{\it flower\/}.  Examples of $(1,3)$-
and $(2,2)$-flowers are shown in Fig.~\ref{g3}.  The DGM net (Fig.~\ref{DGM}) corresponds to the special case of $u=1$
and $v=2$.  All $(u,v)$-flowers are self-similar, as evident from an equivalent method of construction: to produce generation $n+1$, make $w$ copies of the net in generation $n$ and join them at the hubs.

%
It is easy to see, from the second method of construction, that the number of links (the size) of a $(u,v)$-flower of generation $n$ is
\begin{equation}
\label{Mn}
M_n=(u+v)^n= w^n.
\end{equation}
At the same time, the number of nodes (the order) obeys the recursion relation
\[
N_n=wN_{n-1}-w\,,
\]
which, together with the boundary condition $N_1=w$, yields
\begin{equation}
\label{Nn}
N_n=\Big(\frac{w-2}{w-1}\Big)w^n+\Big(\frac{w}{w-1}\Big)\,.
\end{equation}

Similar considerations let us reproduce the full degree distribution.  By construction, $(u,v)$-flowers have only nodes of degree $k=2^m$, $m=1,2,\dots,n$.  Let $N_n(m)$ be the number of nodes of degree $2^m$
in the $(u,v)$-flower of generation $n$, then
\[
N_n(m)=N_{n-1}(m-1)+(w-2)w^{n-1}\delta_{m,1}\,,
\]
leading to
\begin{equation}
\label{Nk}
N_{n}(m) = \left\{ \begin{array}{ll}
(w-2)w^{n-m}, \quad  & m<n,\\
w \quad & m=n.
\end{array} \right.
\end{equation}
As in the DGM case, this corresponds to a scale-free degree distribution, $P(k)\sim k^{-\gamma}$,
of degree exponent
\begin{equation}
\label{gamma}
\gamma=1+\frac{\ln(u+v)}{\ln2}\,.
\end{equation}

\begin{figure}[ht]
  \vspace*{0.3cm}\includegraphics*[width=0.40\textwidth]{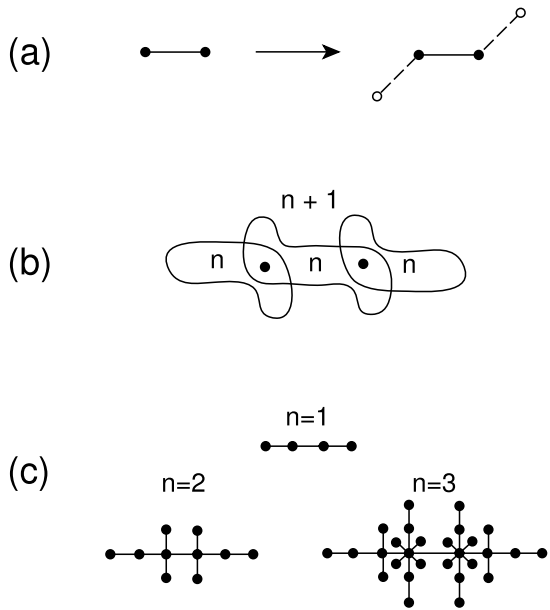}
\caption{The $(1,2)$-tree. 
(a)~Each link in generation $n$ is replaced by a chain of $u=1$ links, to which ends one attaches chains of $v/2=1$ links. (b)~Alternative method of construction highlighting self-similarity: $u+v=3$ replicas of generation $n$ are joined at the hubs.
(c)~Generations $n=1,2,3$.
}
\label{tree}
\end{figure}

Recursive scale-free {\it trees\/} may be defined in analogy to the flower nets.  If $v$ is even, we obtain generation 
$n+1$ of a $(u,v)$-tree by replacing every link in generation $n$ with a chain of $u$ links, and attaching
to each of its endpoints chains of $v/2$ links.  In Fig.~\ref{tree} we show how this works for the $(1,2)$-tree.
If $v$ is odd, we attach to the endpoints (of the chain of $u$ links) chains of length $(v\pm1)/2$.  Different
trees result according to the choices one makes for where to attach the longer (or shorter) chain,
however, they are all similar in their global statistics. Essentially, a $(u,v)$-tree is a $(u,v)$-flower where all the loops are cut open.  The trees may be also constructed by successively joining $w$ replicas at the appropriate hubs, and they too are self-similar.  They share many of the fundamental scaling properties with 
$(u,v)$-flowers: $M_n\sim w^n$, $N_n\sim w^n$, and their degree distribution is scale-free, with 
$\gamma=1+\ln w/\ln 2$.

The self-similarity of $(u,v)$-nets, coupled with the fact that different replicas meet at a {\it single\/} 
node~\cite {ramified}, makes them amenable to exact analysis by renormalization techniques.  The lack of loops, in the case of $(u,v)$-trees, further simplifies their analysis.

\section{Dimensionality}
There is a vast difference between $(u,v)$-nets with $u=1$ and $u>1$.
If $u=1$ the diameter $L_n$ of the $n$-th generation flower (the longest shortest path between any two nodes) scales linearly with $n$.  For example, $L_n=n$ for the $(1,2)$-flower~\cite{dorogovtsev} and
$L_n=2n$ for the $(1,3)$-flower.  It is easy to see that the diameter of the $(1 ,v)$-flower, for $v$ odd, is
$L_n=(v-1)n+(3-v)/2$, and, while deriving a similar result for $v$ even is far from trivial, one can show that $L_n\sim(v-1)n$.

For $u>1$, however, the diameter grows as a power of $n$.  For example, for the $(2,2)$-flower we find
$L_n=2^n$, and, more generally, if $u+v$ is even (and $u>1$),
\[
L_n = \biggl(\frac{u+v}{2} + \frac{v-u}{u-1}\biggr) u^{n-1} - \frac{v-u}{u-1}\,.
\]
For $u+v$ odd, it is harder to obtain $L_n$, but one may establish bounds, 
showing that $L_n\sim u^n$.
To summarize,
\begin{equation}
\label{Ln}
L_n \sim \left\{ \begin{array}{ll}
(v-1)n &  u=1,\\
u^{n} & u > 1,
\end{array} \right.\qquad{\rm flowers.}
\end{equation}
Similar results are quite obvious for the case of $(u,v)$-trees, where \begin{equation}
\label{Ln_trees}
L_n \sim \left\{ \begin{array}{ll}
vn &  u=1,\\
u^{n} & u > 1,
\end{array} \right.\qquad{\rm trees.}
\end{equation}

Since $N_n\sim (u+v)^n$ [Eq.~(\ref{Nn})], we can recast these relations as
\begin{equation}
\label{L}
L \sim \left\{ \begin{array}{ll}
\ln N &  u=1,\\
N^{\ln u/\ln(u+v)} & u > 1.
\end{array} \right.
\end{equation}
Thus, $(u,v)$-nets are {\it small world\/} only in the case of $u=1$.  For $u>1$, the diameter increases as a power of $N$, just as in {\it finite\/}-dimensional objects, and the nets are in fact {\it fractal\/}.  An easy way to see this fractality is as follows.  Given a $(u,v)$-net, we can ``zoom out" (i.e., renormalize) by replacing parallel paths 
of $u$ and $v$ links by a single `super'-link, in a way that reverses the process indicated at the top  of Fig.~\ref{g3}b, say.  This has the effect of rescaling lengths (in {\it chemical\/} space, as measured in number of links  along the shortest path) by a factor of $u$.  At the same time, the number of links (or nodes) in the rescaled net decreases by a factor $(u+v)$.  This mirrors precisely the change of mass in a fractal object
upon the rescaling of length by a factor $b$:
\begin{equation}
\label{defdf}
N(bL)=b^{\df}N(L)\,,
\end{equation}
where $\df$ is the fractal dimension~\cite{chemical}.  In our case,  $N(uL)=(u+v)N(L)$, so
\begin{equation}
\label{df}
\df=\frac{\ln(u+v)}{\ln u}\,,\qquad u>1\,.
\end{equation}
 
\subsection*{Transfinite Fractals}
Small world nets, such as $(1,v)$-nets, are {\it infinite\/}-dimensional.  Indeed, their mass ($N$, or $M$) increases faster than any power (dimension) of their diameter.  Also, note that a naive application of (\ref{df}) to $u\to1$ yields $\df\to\infty$.  We argue that in the case of $(1,v)$-nets one can use their weak self-similarity to define a new measure of dimensionality, $\tdf$, characterizing how mass scales with diameter:
\begin{equation}
\label{deftdf}
N(L+\ell)=e^{\ell\tdf}N(L)\,.
\end{equation}
Instead of a multiplicative rescaling of length, $L\mapsto bL$, we here consider a slower additive mapping,
$L\mapsto L+\ell$, reflecting the small world property.  We term $\tdf$ the {\it transfinite\/} fractal dimension, because it usefully distinguishes between different graphs of infinite dimensionality.  Accordingly, we term 
objects that are self-similar and have infinite dimension (but finite transfinite dimension), such as the $(1,v)$-nets, transfinite fractals, or {\it transfractals\/}, for short.

For $(1,v)$-nets, we see that upon `zooming in' one generation level the mass increases by a factor
of $w=1+v$, while the diameter grows from $L$ to  $L+v-1$ (for flowers), or to $L+v$ (trees).  Hence their transfractal dimension is
\begin{equation}
\label{tdf}
\tdf= \left\{ \begin{array}{ll}
\frac{\ln(1+v)}{v} &(1,v)\text{-trees,}\\ \\
\frac{\ln(1+v)}{v-1} &(1,v)\text{-flowers.}
\end{array} \right.
\end{equation}

There is some arbitrariness in the selection of $e$ as the base of the exponential in the definition (\ref{deftdf}), that we are unable
to remove at this time.  We note, however, that the base is inconsequential for the sake of
comparison between dimensionalities of different objects.  Also, {\it scaling relations\/} between various transfinite exponents
hold, irrespective of the choice of base.   As an illustration of this fact, consider the scaling relation
\begin{equation}
\label{gfrac}
\gamma=1+\frac{\df}{\dk}\,,
\end{equation}
valid for fractal scale-free nets of degree exponent $\gamma$~\cite{Song,Song2}.  $\dk$ is an exponent characterizing the self-similarity of the net with regards to its degree distribution: suppose that renormalization carries clusters of links
of diameter $b$  into a single `super'-link (of length one), then the new degree distribution, $P'$, is related to
the old distribution via
\begin{equation}
\label{scalePk}
P'(k)=b^{\dk}P(b^{-dk}k)\,.
\end{equation}
For example, in the fractal $(u,v)$-nets (with $u>1$) renormalization reduces lengths by a factor $b=u$ and all degrees are reduced by a factor of 2, so $b^{\dk}=2$. Thus $\dk=\ln2/\ln u$, and since $\df=\ln(u+v)/\ln u$ and $\gamma=1+\ln(u+v)/\ln2$, as discussed above, 
the relation (\ref{gfrac}) is indeed satisfied.

For transfractals,  renormalization reduces distances by an {\it additive\/}
length, $\ell$, and we express the self-similarity manifest in the degree distribution as
\begin{equation}
\label{scaletPk}
P'(k)=e^{\ell\tdk}P(e^{-\ell\tdk}k)\,,
\end{equation}
where $\tdk$ is the transfinite exponent analogous to $\dk$.  Renormalization of the transfractal $(1,v)$-nets
reduces the link lengths by $\ell=v-1$ (for flowers), or $\ell=v$ (trees), while all degrees are halved.
Thus,
\[
 \tdk= \left\{ \begin{array}{ll}
\frac{\ln2}{v} &(1,v)\text{-trees,}\\ \\
\frac{\ln2}{v-1} &(1,v)\text{-flowers.}
\end{array} \right.
\]
Along with (\ref{tdf}), this result confirms that the scaling relation  
\begin{equation}
\label{gtfrac}
\gamma=1+\frac{\tdf}{\tdk}
\end{equation}
is valid also for transfractals, and regardless of the choice of base.
A general proof of this relation is practically identical to the proof of~(\ref{gfrac})~\cite{Song}, merely replacing fractal with transfractal scaling throughout the argument.  

Before closing this section, let us illustrate a practical use of dimensionalities and scaling relations.  For fractal scale-free nets  the size, $N$, and the nets' highest degree, $K$, scale with the diameter, $L$, as:
\[
N\sim L^{\df}\,,\qquad K\sim L^{\dk}.
\]
Imagine, indeed, starting with a net of size $N$ and diameter $L$ and renormalizing $n$ times, until the diameter and size shrink to order one.  Clearly, $L\sim b^n$, and $N\sim (b^{\df})^n$ [see Eq.~(\ref{defdf})],
leading to the first relation.  At the same time $K$ renormalizes 
to order one as well, and, using (\ref{scalePk}), we conclude that $K\sim(b^{\dk})^n$, confirming the second
relation.  Putting the two together, we find that
$K\sim L^{\dk}\sim N^{\dk/\df}$.  Thus, in view of (\ref{gfrac}), we obtain
\[
K\sim N^{\frac{1}{\gamma-1}}\,,
\]
a useful result that has been derived elsewhere, for random scale-free nets,  by independent means~\cite{cohen}.

For scale-free transfractals, following $m=L/\ell$ renormalizations
the diameter and mass reduce to order one, and the scaling (\ref{deftdf}) implies $L\sim m\ell$, $N\sim e^{m\ell\tdf}$, so that
\[
L\sim \frac{1}{\tdf}\ln N\,,
\]
in accordance with their small world property.  At the same time the scaling (\ref{scaletPk}) implies
$K\sim e^{m\ell\tdk}$, or $K\sim N^{\tdk/\tdf}$.  Using the scaling relation (\ref{gtfrac}), we rederive 
$K\sim N^{1/(\gamma-1)}$, which is indeed valid for scale-free nets {\it in general\/}, be they fractal or transfractal.

\section{MULTISCALING}
Inherent in scale-free nets is a lack of translational symmetry with regards to the properties of low-degree and higher-degree nodes.  As a simple example, the typical distance between nodes in the DGM network of
generation $n$ increases linearly with $n$, whereas the distance between the highest connected nodes 
is just one.  In this section, we focus on the different scalings for {\it resistance\/} and  first passage time for {\it diffusion\/} between {\it hubs\/} versus between regular {\it nodes\/}.  Remarkably, despite the different scalings, the Einstein relation connecting the two phenomena holds separately (for hubs and nodes), both for fractal and transfractal $(u,v)$-nets.

\subsection*{Scaling and the Einstein Relation}
Suppose that each link in a graph of diameter $L$ and size $N$ has one unit of resistance.  If the graph is fractal, the typical resistance between any two sites, $R$, scales as a power law:
\[
R\sim L^{\zeta}\,,
\]
where $\zeta$ is the resistance exponent (in chemical space).  Consider also diffusion, or random walks, on the graph, where at each time step the walker hops from its current location to one of the neighboring nodes, with equal probability.  The characteristic time for diffusion between any two nodes  scales too as
a power law:
\[
T\sim L^{\dw}\,,
\]
where $\dw$ is the walk dimension (again, in chemical space).  The Einstein relation for resistance and diffusion states that 
\begin{equation}
\label{Einstein}
\zeta=\dw-\df\,,
\end{equation}
or, equivalently,  $R\sim T/N$.

For transfractals, the exponents $\zeta$, $\dw$, and $\df$ are infinite.  Instead, we have 
$R(L+\ell)\sim e^{\ell\tzeta}R(L)$, $T(L+\ell)\sim e^{\ell\tdw}T(L)$, and $N(L+\ell)\sim e^{\ell\tdf}N(L)$. The scaling $R\sim T/N$ suggests that
the Einstein relation is then valid also for the transfinite exponents:
\begin{equation}
\label{tEinstein}
\tzeta=\tdw-\tdf\,.
\end{equation}
(Notice that the choice for  base in the definition of the transfinite exponents is inconsequential.)
In what follows, we shall see that this is indeed the case, at least for $(u,v)$-nets.  Moreover, we shall also demonstrate that while {\it different\/} sets of $\zeta$ and $\dw$ (and $\tzeta$ and $\tdw$)  exponents characterize hubs and nodes, the Einstein relation is obeyed in all instances.

\subsection*{Hubs and Nodes}
Imagine a $(u,v)$-net iterated {\it ad infinitum\/}.  In this case we may distinguish between two types of nodes, according to their degree:
(a)~Nodes whose degree is infinite, which we term {\it hubs\/}, and (b)~nodes of finite degree, which, in lack of a better name, we shall term {\it nodes\/}.  Stated differently, nodes are the sites that had been introduced during the last $m$ iterations, even as the total number of iterations $n\to\infty$.  Perhaps counter-intuitively,
the fraction of nodes is   
\[
\lim_{n\to\infty}\frac{\sum_{l=m+1}^{\infty} N_n(l)}{\sum_{l=1}^{\infty} N_n(l)}=1\,,
\]
despite the fact that $m$ is kept finite in the limit.  [One can see this by using $N_n(l)$ of Eq.~(\ref{Nk}).]  
Thus almost all sites are nodes, in a statistical sense, lending some justification to our awkward terminology.
Although hubs have zero measure they dominate many of the more exotic properties of scale-free nets,
such as resilience to random dilution~\cite{cohen,stauffer}, and they may not be ignored.

\subsection*{Scaling for Hubs}
We wish to derive scaling relations for transport between hubs.  A practical way to pick hubs is thinking of a net of generation $m$ (a finite integer) which is then further iterated {\it ad infinitum\/}.  All the sites present at generation $m$
will thus become hubs.

Consider first resistance between hubs (the nodes of generation $m$) for $(u,v)$-trees.  Each
iteration of the tree results in an increase of the resistance between hubs by a factor of $u$  (the dangling
chains do not affect the resistance).  Thus, $R(n)=u^{n-m}R(m)$.  Taking the limit of $n\to\infty$ while 
keeping $m$ fixed we conclude that
\begin{equation}
\label{eq17}
R_{\rm hubs}(n)\sim u^n\,;\qquad\text{trees.}
\end{equation}
This relation suggests that for transfractal trees ($u=1$) the resistance between hubs remains constant.
This is indeed the case: since the hubs could only be introduced at a finite generation, the distance between any two hubs is finite, because successive iterations do not change the distance between existing nodes when $u=1$.

For $(u,v)$-flowers, each iteration replaces a link of resistance $1$ with two parallel chains of resistance 
$u$ and $v$, of a combined resistance of $uv/(u+v)$.  Thus,
\begin{equation}
\label{eq18}
R_{\rm hubs}(n)\sim \Big(\frac{uv}{u+v}\Big)^n\,;\qquad\text{flowers.}
\end{equation}  
$R$ decreases with size for transfractal flowers ($u>1$) and increases for fractals; the $(2,2)$-flower is a marginal case where the resistance between hubs remains constant upon rescaling~\cite{2d}.

\begin{figure}[ht]
  \vspace*{0.3cm}\includegraphics*[width=0.40\textwidth]{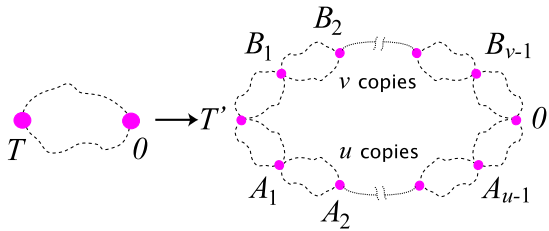}
\caption{
Rescaling of FPT for $(u,v)$-flowers.  The FPT, $T$, between two hubs in generation $n$ (left) becomes
$T'$ in generation $n+1$ (right).  The $A_i$ and $B_j$ are FPTs from intermediate nodes
in generation $n+1$ to the target hub~$O$.}
\label{fpt2}
\end{figure}

Next we analyze the first passage time (FPT) --- the average time needed for a walker to reach from one hub to another for the first time~\cite{bollt,fptbook}.  Assume that in the $(u,v)$-flower of generation $n$ the FPT  is $T$, and compute $T'$,  the FPT between the same two hubs in generation $n+1$ (Fig.~\ref{fpt2}).  Let $A_i$ ($i=1,2,\dots,u-1$) and $B_j$ ($j=1,2,\dots,
v-1$) be the first passage times to the target hub from the intervening nodes in generation $l+1$.  The various FPTs obey the backward equations
\[
\begin{split}
&T'=\frac{1}{2}(T+A_1)+\frac{1}{2}(T+B_1)\,,\\
&A_1=\frac{1}{2}(T+T')+\frac{1}{2}(T+A_2)\,,\\
&A_2=\frac{1}{2}(T+A_1)+\frac{1}{2}(T+A_3)\,,\\
&\>\>\>\>\>\>\>\>\vdots\\
&A_{u-1}=\frac{1}{2}(T+A_{u-2})+\frac{1}{2}T\,,\\
&B_1=\frac{1}{2}(T+T')+\frac{1}{2}(T+B_2)\,,\\
&B_2=\frac{1}{2}(T+B_1)+\frac{1}{2}(T+B_3)\,,\\
&\>\>\>\>\>\>\>\>\vdots\\
&B_{v-1}=\frac{1}{2}(T+B_{v-2})+\frac{1}{2}T\,.
\end{split}
\]
Eliminating the $A_i$ and $B_j$ we find $T'=(uv)T$.  Thus, in $(u,v)$-flowers the FPT between hubs scales
as
\begin{equation}
\label{eq19}
T_{\rm hubs}(n)\sim (uv)^n\,;\qquad\text{flowers}.
\end{equation}
A similar analysis for $(u,v)$-trees with $v$ even yields 
\begin{equation}
\label{eq20}
T_{\rm hubs}(n)\sim[u (u+v)]^n\,;\qquad\text{trees}.
\end{equation}
The case of odd $v$ is more involved, to cover all possible trees that result from our iteration rules, however, the results are not different from the one for $v$ even in any significant way.

We can now compute the various exponents pertaining to hubs.  For example, for fractal flowers ($u>1$)
upon zooming in
the resistance increases by $(uv)/(u+v)$ and the FPT by $(uv)$, while lengths increase by $b=u$, hence
$\zeta=\ln[uv/(u+v]/\ln u$ and $\dw=\ln(uv)/\ln u$.  For transfractal flowers resistance decreases by $v/(1+v)$
and FPT increases by a factor $v$, while lengths increase additively by $v-1$, hence
$\tzeta=\ln[v/(1+v)]/(v-1)$ and $\tdw=(\ln v)/(v-1)$.  Since for flowers $\df=\ln(u+v)/\ln u$ and $\tdf=\ln(1+v)/(v-1)$, we confirm that the Einstein relations (\ref{Einstein}) and (\ref{tEinstein}) work for this case.
Note, however, that it is easier to check the scaling $R\sim T/N$ directly, without bothering with the exponents.

\subsection*{Scaling for Nodes}
\noindent{\bf Trees:}

Consider first the scaling of transport for nodes in $(u,v)$-trees.  The resistance between two nodes is the same as their distance in chemical space (the dangling chains do not affect resistance), and since the average distance between nodes is of order $L$, we have
\begin{equation}
\label{eq21}
R_{\rm nodes}(n)\sim \left\{ \begin{array}{ll}
vn &u=1,\\ 
u^n &u>1.
\end{array} \right.
\end{equation}
The situation is most interesting for transfractal trees ($u=1$), where the scaling $R_{\rm nodes}\sim
vn$ is different from $R_{\rm hubs}\sim\text{const.}$, of Eq~(\ref{eq17}).  The scaling for the FPT between nodes in $(1,v)$-trees is also different than for hubs.  From the backward equations one can show that the average FPT between nodes scales as 
\begin{equation}
T_{\rm nodes}(n)\sim n(1+v)^n\,,\qquad\text{$(1,v)$-trees,}
\end{equation}
as opposed to $T_{\rm hubs}\sim(1+v)^n$ of Eq.~(\ref{eq20}).
Remarkably, the Einstein relation  $R\sim T/N$ holds also for nodes, since the extra factors of $n$ in the scalings of $R_{\rm nodes}$ and $T_{\rm nodes}$ cancel out (there is only one global scaling form for $N$)~\cite{correction}.

For fractal trees ($u>1$), the scaling $R_{\rm nodes}\sim u^n$ is the same as for hubs.  Using the backward equations one can show that  $T_{\rm nodes}(n)\sim [u(u+v)]^n$, which is also the same as for hubs.  Thus, for fractal trees there is no difference in scaling for hubs and nodes and the networks seem homogeneous in this sense, despite the great spread in the degrees of the nodes.

\medskip\noindent
{\bf Flowers:}

The scaling of transport between nodes and between hubs is most dramatically different in transfractal $(1,v)$-flowers.
In \cite{bollt} it was shown that the FPT between nodes of the DGM network scales as
\begin{equation}
T_{\rm nodes}(n)\sim 3^n,\qquad\text{$(1,2)$-flower.}
\end{equation}
This is functionally slower than for hubs, $T_{\rm hubs}\sim 2^n$, of Eq.~(\ref{eq19}).

Consider now the scaling for resistance.  For hubs, we have seen that $R_{\rm hubs}\sim(2/3)^n$ [Eq.~(\ref{eq18})], so that the resistance vanishes as $n\to\infty$.  We argue that, in contrast, the resistance between 
nodes tends to a constant  (as $n\to\infty$). It is sufficient to focus on nodes of degree 2, which
according to the distribution~(\ref{Nk}) constitute about two-thirds of all nodes.  A {\it lower bound\/} to the
resistance between nodes of degree 2 may be obtained as follows:   Assign to each link  opposite a 2-degree node resistance zero, effectively short-circuiting the two links.  Following this transformation the $(1,2)$-flower becomes a star graph, where each of the 2-degree nodes is connected to a central hub through two parallel links (of combined resistance $1/2$). Thus, the resistance between nodes of degree 2 in the flower is greater than 1.  It is easy to convince oneself that the resistance between 2-nodes is also  bounded from above  (linear increases in the distance between nodes is more than compensated by the exponential decay in the resistance of intervening subgraphs).
To summarize,
\begin{equation}
R_{\rm nodes}(n)\sim \text{const.},\qquad\text{$(1,2)$-flower,}
\end{equation}
as $n\to\infty$.

The radically different scaling for hubs and nodes, in the $(1,2)$-flower, implies that in this case one cannot even speak of a global resistance exponent $\tzeta$, or a global walk exponent $\tdw$, but rather 
define exponents for various subsets of the graph.  Nevertheless, the Einstein relation is satisfied separately for the different subsets:  For hubs, $R_{\rm hubs}\sim(2/3)^n$, $T_{\rm hubs}\sim2^n$, and
$N\sim 3^n$, so $R\sim T/N$ is satisfied.  For nodes, $R_{\rm nodes}\sim{\rm const.}$,
$T_{\rm nodes}\sim3^n$ (and $N\sim3^n$), and the Einstein relation is satisfied once again. 

\begin{figure}[ht]
  \vspace*{0.3cm}\includegraphics*[width=0.35\textwidth]{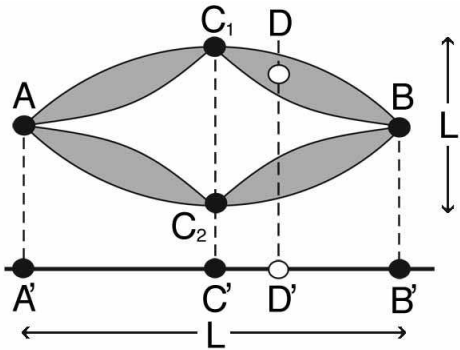}
\caption{Projection of the $(2,2)$-flower (top) on the line (bottom).  All sites equidistant from the hub $A$ are projected onto a single point on the line.  Hubs are denoted by  full circles ($\bullet$).  Nodes ($\circ$) reside in the gray-shaded regions, in this schematic representation.  There are typically $L\sim\sqrt{N}$
nodes along a projection line such as $DD'$, but only one is shown, for clarity.}
\label{flower1d}
\end{figure}

Fractal $(u,v)$-flowers are similar to fractal $(u,v)$-trees, in that the scaling of resistance and FPT for nodes is essentially the same as for hubs.  Consider, for example, the $(2,2)$-flower,  where we are able to analyze diffusion heuristically, as follows.  We map the flower onto a one-dimensional chain by associating all sites that are equidistant from a hub, $A$, with a single point on the line (and at the same distance), as depicted schematically in Fig.~\ref{flower1d}.  Note that all sites projected onto a single point  have an equal number of links emanating to the left and right (apart from the hubs $A$ and $B$), so a random walk on the flower appears  as a non-biased random walk on the projected line.  

The length of the projection equals the distance between the hubs $A$ and $B$, $L=2^n\sim \sqrt{N}$ (since 
$N\sim4^n$).  It follows that there are typically $\sqrt{N}$ nodes along a projection line such as $DD'$.  (However, the number of hubs along a projection line such as $CC'$ remains constant as $N\to\infty$.)
Consider diffusion between  two hubs such as $A$ and $C_1$.  The distance between the projected hubs, 
$A'$ and $C'$, is $\sim L$, so it takes $\sim L^2$ steps to diffuse from $A'$ to $C'$.  Suppose that there are $m$ hubs in the projection line $CC'$, then the probability that arrival at $C'$ coincides with arrival to $C_1$
in the flower is $1/m$.  That is, the walker needs to return to $CC'$ about $m$ times to hit
$C_1$ with a significant probability.  Since the FPT for each return takes $\sim L$ steps, the total FPT
is expected to scale as $T_{\rm hubs}\sim L^2+mL\sim L^2\sim N$, in agreement with Eq.~(\ref{eq19}).
Consider now diffusion from some arbitrary node to node $D$, say.  In this case $m\sim L$ (which diverges as $N\to\infty$), but the final result is the same: $T_{\rm nodes}\sim L^2+mL\sim L^2\sim N$.

The scaling of resistance for hubs and nodes is also the same; $R_{\rm hubs}\sim R_{\rm nodes}\sim{\rm const}$.  One can see that this is true for hubs from Eq.~(\ref{eq18}),  and for nodes, by systematically replacing parallel chains of two links by
a single link (of equivalent resistance $1$). 

It is somewhat surprising that such dramatically different subsets as hubs and nodes exhibit the same scaling
for resistance and FPT in fractal $(u,v)$-trees and flowers.  For the theorist, it is also a desirable property, for it makes then sense characterizing all sites by a common global average.  Nevertheless, there remain strong
asymmetries between hubs and nodes with regards to other physical attributes (besides the degree and the scaling exponents).  Consider, for example, the question of recurrence. We term a site {\it recurrent\/} if a random walker returns to it almost surely, as $t\to\infty$, in the limit of $N\to\infty$.  Otherwise, the site is {\it transient\/}.  It is well known that all sites on the line are recurrent.  It then follows that hubs of the $(2,2)$-flower are recurrent (since each recurrence on the line hits the hub in question with probability $1/m>0$), but nodes are transient (since $1/m\sim1/\sqrt{N}\to0$, as $N\to\infty$).  The same is true with respect to nodes and hubs of fractal trees.

\section{Discussion}

In summary, we have introduced a class of recursive scale-free nets, $(u,v)$-flowers and trees, that yield themselves to exact analysis.  All networks in this family are self-similar, in the sense that each net contains $u+v$ subgraphs that resemble the whole.  For $u=1$ ($u\leq v$), the networks are small-world: their diameter, $L$, increases logarithmically with their size, $L\sim \ln N$.  For $u>1$, the diameter increases
as a power of $N$, so the networks are not small-world, but owing to this scaling they possess well-defined
dimensionalities characteristic of fractals (in chemical space).  
For $u=1$, the nets are infinite-dimensional.  Exploiting their self-similarity we were able to define {\it transfinite\/}
dimensions, that usefully characterize the nets while taking into account their small-world scaling.

An especially useful example is provided by comparing the $(2,2)$- to the $(1,3)$-flower.  In both cases $\gamma=3$
(their degree distributions are {\it identical\/}, node for node),
but the former is a fractal, while the latter is a transfractal, and there are vast differences in the scaling of
resistance, and diffusion, as analyzed in the text.  Another amusing difference concerns their degree of assortativity (the extent to which nodes of similar degree connect with one 
another)~\cite{newman1,newman2}.  In the transfractal $(1,3)$-flower, nodes of degree $2^m$ and $2^{m+1}$ are only {\it one\/} link apart, 
and the assortativity index is 0; while in the fractal $(2,2)$-flower the same nodes are $2^{m-1}$ links apart, and
its assortativity index tends to $-1/2$ (as $N\to\infty$)~\cite{hernan}, indicating a high degree of disassortativity.  This
is curiously in line with what is found in naturally occurring fractal and non-fractal nets~\cite{Song2,strogatz}.  Further study of the $(2,2)$-
and $(1,3)$-flowers will undoubtedly uncover other interesting differences.

We have also addressed the absence of self-averaging in scale-free nets, due to the wide distribution of degrees.
Making the distinction between {\it hubs\/} (nodes whose degree is infinite, as $N\to\infty$) and {\it nodes\/} (whose degrees remains finite), we showed that they satisfy different scaling laws, characterized by different transfinite exponents, in the case of transfractal ($u=1$) nets.  Nevertheless, the Einstein relation for resistance and diffusion
is obeyed separately for the two subsets, despite the different scalings.  In the fractal nets ($u>1$) we found the same scaling for hubs and nodes, however, the two subsets still differ: for example,  hubs are recurrent (walks starting from a hub return to it almost surely, even as $N\to\infty$), whereas nodes are not.

There remain several interesting open questions.  Exact recursive nets merit further study, as they offer much insight into stochastic scale-free graphs.  An important question is to what extent the self-similarity of our recursive models is present in stochastic scale-free graphs, and whether the stochastic nets could be characterized by transfinite dimensions.  Random nets with $2<\gamma<3$ are {\it ultra\/}-small-world:
their diameter scales as $L\sim\ln\ln N$~\cite{cohen2}.  It would be useful to invent recursive models that exhibit this scaling, and study their properties.
We anticipate that ultra-small-world nets would have diverging transfinite dimensions, and that one could define
dimensions of {\it higher\/} transfinite order that usefully characterize them, in analogy to what was done in the present work.  

The question of multiscaling in scale-free nets will be the subject of future research.  The gross distinction made here between hubs and nodes should be refined, to capture the full spectrum of different scalings of various nodes subsets, in the spirit of what was done with multifractals.  What are the general conditions required for  ``detailed scaling" --- scaling relations between exponents that hold separately for the various subsets of nodes?  Would ``detailed scaling" be found also in stochastic nets?  Would analogous relations be valid for ultra-small world scale-free nets? ---After all, the relation $K\sim N^{1/(\gamma-1)}$ holds also in their case.

Recently, Song {\it et al\/}., have studied naturally occurring scale-free nets that seem to be fractal and small-world at the same time~\cite{Song,Song2}.  A possible model for such behavior is achieved by mixing the construction rules for fractal and transfractal recursive nets (with the same value of $u+v$, or the degree exponent $\gamma$).  Suppose, for example, that we build a $(2,2)$-flower up to generation $m$, and
thereafter we string replicas together according to the rule for $(1,3)$-flowers, up to generation $n\,\,(>m)$.  The resulting net, of $\gamma=3$, is fractal
up to distances $L_*\sim 2^m$.  However, the subsequent scaling is small-world: $L\sim2(n-m)L_*$, according to Eq.~(\ref{Ln}), or $L\sim\ln N$  (since $N\sim4^n$, and $m$ remains finite as $n\to\infty$).  Such models would be an asset to the study of natural networks, where the various scalings are hard to pin down due to their modest sizes.

\acknowledgments
We thank James P. Bagrow for numerous discussions.
We also thank the NSF (PHY0555312), ONR, Israel Science Foundation, European NEST project DYSONET,
FONCyT (PICT-O2004/370),  and the Israeli Center for Complexity Science for
financial support.



\begin{thebibliography}{99}
\bibitem{correction}
Note that the extra factors of $n$ in the scaling for nodes would not show up in the exponents $\tzeta$
and $\tdw$, and may in fact be regarded as {\it corrections to scaling\/}.  It is reassuring that the functional form of the Einstein relation, $R\sim T/N$, works despite the corrections!
\end{thebibliography}
\end{document}